\documentclass[10pt,a4paper,onecolumn]{article}
\usepackage{marginnote}
\usepackage{graphicx}
\usepackage{xcolor}
\usepackage{authblk,etoolbox}
\usepackage{titlesec}
\usepackage{calc}
\usepackage{tikz}
\usepackage{hyperref}
\hypersetup{colorlinks,breaklinks,
            urlcolor=[rgb]{0.0, 0.5, 1.0},
            linkcolor=[rgb]{0.0, 0.5, 1.0}}
\usepackage{caption}
\usepackage{tcolorbox}
\usepackage{amssymb,amsmath}
\usepackage{ifxetex,ifluatex}
\usepackage{seqsplit}
\usepackage{xstring}

\usepackage{float}
\let\origfigure\figure
\let\endorigfigure\endfigure
\renewenvironment{figure}[1][2] {
    \expandafter\origfigure\expandafter[H]
} {
    \endorigfigure
}

\usepackage{fixltx2e} % provides \textsubscript
\usepackage[
  backend=biber,
]{biblatex}
\bibliography{paper.bib}

\let\textttOrig=\texttt
\def\texttt#1{\expandafter\textttOrig{\seqsplit{#1}}}
\renewcommand{\seqinsert}{\ifmmode
  \allowbreak
  \else\penalty6000\hspace{0pt plus 0.02em}\fi}

\makeatletter
\let\href@Orig=\href
\def\href@Urllike#1#2{\href@Orig{#1}{\begingroup
    \def\Url@String{#2}\Url@FormatString
    \endgroup}}
\def\href@Notdoi#1#2{\def\tempa{#1}\def\tempb{#2}%
  \ifx\tempa\tempb\relax\href@Urllike{#1}{#2}\else
  \href@Orig{#1}{#2}\fi}
\def\href#1#2{%
  \IfBeginWith{#1}{https://doi.org}%
  {\href@Urllike{#1}{#2}}{\href@Notdoi{#1}{#2}}}
\makeatother

\usepackage[top=3.5cm, bottom=3cm, right=1.5cm, left=1.0cm,
            headheight=2.2cm, reversemp, includemp, marginparwidth=4.5cm]{geometry}

\titleformat{\section}
  {\normalfont\sffamily\Large\bfseries}
  {}{0pt}{}
\titleformat{\subsection}
  {\normalfont\sffamily\large\bfseries}
  {}{0pt}{}
\titleformat{\subsubsection}
  {\normalfont\sffamily\bfseries}
  {}{0pt}{}
\titleformat*{\paragraph}
  {\sffamily\normalsize}

\usepackage{fancyhdr}
\makeatletter
\let\ps@plain\ps@fancy
\fancyheadoffset[L]{4.5cm}
\fancyfootoffset[L]{4.5cm}

\definecolor{linky}{rgb}{0.0, 0.5, 1.0}

\newtcolorbox{repobox}
   {colback=red, colframe=red!75!black,
     boxrule=0.5pt, arc=2pt, left=6pt, right=6pt, top=3pt, bottom=3pt}

\newcommand{\ExternalLink}{%
   \tikz[x=1.2ex, y=1.2ex, baseline=-0.05ex]{%
       \begin{scope}[x=1ex, y=1ex]
           \clip (-0.1,-0.1)
               --++ (-0, 1.2)
               --++ (0.6, 0)
               --++ (0, -0.6)
               --++ (0.6, 0)
               --++ (0, -1);
           \path[draw,
               line width = 0.5,
               rounded corners=0.5]
               (0,0) rectangle (1,1);
       \end{scope}
       \path[draw, line width = 0.5] (0.5, 0.5)
           -- (1, 1);
       \path[draw, line width = 0.5] (0.6, 1)
           -- (1, 1) -- (1, 0.6);
       }
   }

\patchcmd{\@maketitle}{center}{flushleft}{}{}
\patchcmd{\@maketitle}{center}{flushleft}{}{}
\patchcmd{\@maketitle}{\LARGE}{\LARGE\sffamily}{}{}
\def\maketitle{{%
  
  \AB@maketitle}}
\makeatletter
\renewcommand\AB@affilsepx{ \protect\Affilfont}
\renewcommand\AB@affilnote[1]{{\bfseries #1}\hspace{3pt}}
\renewcommand{\affil}[2][]%
   {\newaffiltrue\let\AB@blk@and\AB@pand
      \if\relax#1\relax\def\AB@note{\AB@thenote}\else\def\AB@note{#1}%
        \setcounter{Maxaffil}{0}\fi
        \begingroup
        \let\href=\href@Orig
        \let\texttt=\textttOrig
        \let\protect\@unexpandable@protect
        \def\thanks{\protect\thanks}\def\footnote{\protect\footnote}%
        \@temptokena=\expandafter{\AB@authors}%
        {\def\\{\protect\\\protect\Affilfont}\xdef\AB@temp{#2}}%
         \xdef\AB@authors{\the\@temptokena\AB@las\AB@au@str
         \protect\\[\affilsep]\protect\Affilfont\AB@temp}%
         \gdef\AB@las{}\gdef\AB@au@str{}%
        {\def\\{, \ignorespaces}\xdef\AB@temp{#2}}%
        \@temptokena=\expandafter{\AB@affillist}%
        \xdef\AB@affillist{\the\@temptokena \AB@affilsep
          \AB@affilnote{\AB@note}\protect\Affilfont\AB@temp}%
      \endgroup
       \let\AB@affilsep\AB@affilsepx
}
\makeatother

\renewcommand\Affilfont{\sffamily\small\mdseries}
\ifnum 0\ifxetex 1\fi\ifluatex 1\fi=0 % if pdftex
  \usepackage[T1]{fontenc}
  \usepackage[utf8]{inputenc}

\else % if luatex or xelatex
  \ifxetex
    \usepackage{mathspec}
  \else
    \usepackage{fontspec}
  \fi
  \defaultfontfeatures{Ligatures=TeX,Scale=MatchLowercase}

\fi
\IfFileExists{upquote.sty}{\usepackage{upquote}}{}
\IfFileExists{microtype.sty}{%
\usepackage{microtype}
\UseMicrotypeSet[protrusion]{basicmath} % disable protrusion for tt fonts
}{}

\usepackage{hyperref}
\hypersetup{unicode=true,
            pdftitle={deeplenstronomy: A dataset simulation package for strong gravitational lensing},
            pdfborder={0 0 0},
            breaklinks=true}
\let\addcontentslineOrig=\addcontentsline
\def\addcontentsline#1#2#3{\bgroup
  \let\texttt=\textttOrig\addcontentslineOrig{#1}{#2}{#3}\egroup}
\let\markbothOrig\markboth
\def\markboth#1#2{\bgroup
  \let\texttt=\textttOrig\markbothOrig{#1}{#2}\egroup}
\let\markrightOrig\markright
\def\markright#1{\bgroup
  \let\texttt=\textttOrig\markrightOrig{#1}\egroup}

\usepackage{graphicx,grffile}
\makeatletter
\def\maxwidth{\ifdim\Gin@nat@width>\linewidth\linewidth\else\Gin@nat@width\fi}
\def\maxheight{\ifdim\Gin@nat@height>\textheight\textheight\else\Gin@nat@height\fi}
\makeatother
\setkeys{Gin}{width=\maxwidth,height=\maxheight,keepaspectratio}
\IfFileExists{parskip.sty}{%
\usepackage{parskip}
}{% else
\setlength{\parindent}{0pt}
\setlength{\parskip}{6pt plus 2pt minus 1pt}
}
\ifx\paragraph\undefined\else
\let\oldparagraph\paragraph
\renewcommand{\paragraph}[1]{\oldparagraph{#1}\mbox{}}
\fi
\ifx\subparagraph\undefined\else
\let\oldsubparagraph\subparagraph
\renewcommand{\subparagraph}[1]{\oldsubparagraph{#1}\mbox{}}
\fi

\title{lenstronomy II: A gravitational lensing software ecosystem}

        \author[1, 2]{Simon Birrer\footnote{Corresponding author: sibirrer@stanford.edu}}
          \author[3, 4]{Anowar J.~Shajib}
          \author[5]{Daniel Gilman}
          \author[6]{Aymeric Galan}
          \author[1, 2]{Jelle Aalbers}
          \author[6]{Martin Millon}
          \author[7, 8]{Robert Morgan}
          \author[9]{Giulia Pagano}
          \author[1, 2]{Ji~Won Park}
          \author[10]{Luca Teodori}
          \author[11]{Nicolas Tessore}
          \author[1]{Madison Ueland}
          \author[12]{Lyne Van~de~Vyvere}
          \author[1, 2]{Sebastian Wagner-Carena}
          \author[13]{Ewoud Wempe}
          \author[14]{Lilan Yang}
          \author[15]{Xuheng Ding}
          \author[4]{Thomas Schmidt}
          \author[12]{Dominique Sluse}
          \author[16]{Ming Zhang}
          \author[17]{Adam Amara}

      \affil[1]{Kavli Institute for Particle Astrophysics and Cosmology and Department of Physics, Stanford University, Stanford, CA 94305, USA}
      \affil[2]{SLAC National Accelerator Laboratory, Menlo Park, CA, 94025, USA}
      \affil[3]{Department of Astronomy \& Astrophysics, University of Chicago, Chicago, IL 60637, USA}
      \affil[4]{Department of Physics and Astronomy, University of California, Los Angeles, CA 90095, USA}
      \affil[5]{Department of Astronomy and Astrophysics, University of Toronto, 50 St. George Street, Toronto, ON, M5S 3H4, Canada}
      \affil[6]{Institute of Physics, Laboratory of Astrophysics, Ecole Polytechnique F\'ed\'erale de Lausanne (EPFL), Switzerland}
      \affil[7]{Physics Department, University of Wisconsin-Madison, 1150 University Avenue Madison, WI  53706, USA}
      \affil[8]{Legacy Survey of Space and Time Corporation Data Science Fellowship Program, USA}
      \affil[9]{Independent Researcher}
      \affil[10]{Weizmann Institute, 234 Herzl Street, Rehovot, 7610001 Israel}
      \affil[11]{Department of Physics and Astronomy, University College London, Gower Street, London, WC1E 6BT, UK}
      \affil[12]{STAR Institute, Universit\'e de Liège, Quartier Agora - All\'ee du six Août, 19c, B-4000 Liège, Belgium}
      \affil[13]{Kapteyn Astronomical Institute, University of Groningen, PO Box 800, 9700 AV Groningen, the Netherlands}
      \affil[14]{School of Physics and Technology, Wuhan University, Wuhan 430072, China}
      \affil[15]{Kavli IPMU (WPI), UTIAS, The University of Tokyo, Kashiwa, Chiba 277-8583, Japan}
      \affil[16]{Xinjiang Astronomical Observatory, Chinese Academy of Sciences, 150 Science 1-Street, Urumqi 831001, China}
      \affil[17]{Institute of Cosmology and Gravitation, University of Portsmouth, Portsmouth PO1 3FX, UK}
  
\date{\vspace{-5ex}}

\begin{document}
\maketitle

\marginpar{
  \sffamily\small

  {\bfseries DOI:} \href{https://doi.org/10.21105/joss.03283}{\color{linky}{10.21105/joss.03283}}

  \vspace{2mm}

  {\bfseries Software}
  \begin{itemize}
    \setlength\itemsep{0em}
    \item \href{https://github.com/openjournals/joss-reviews/issues/3283}{\color{linky}{Review}} \ExternalLink
    \item \href{https://github.com/sibirrer/lenstronomy}{\color{linky}{Repository}} \ExternalLink
    \item \href{https://zenodo.org/record/4913700}{\color{linky}{Archive}} \ExternalLink
  \end{itemize}

  \vspace{2mm}

  {\bfseries Submitted:} 03 May 2021\\
  {\bfseries Published:} 08 June 2021

  \vspace{2mm}
  {\bfseries License}\\
  Authors of papers retain copyright and release the work under a Creative Commons Attribution 4.0 International License (\href{https://creativecommons.org/licenses/by/4.0/}{\color{linky}{CC BY 4.0}}).
}

\section{Summary}\label{summary}

\textsc{lenstronomy} is an Astropy-affiliated (\hyperlink{astropy:2013}{Astropy Collaboration 2013}, \hyperlink{astropy:2018}{2018}) Python package for gravitational lensing simulations and analyses.
\textsc{lenstronomy} was introduced by \hyperlink{lenstronomy1}{Birrer \& Amara (2018)} and is based on the linear basis set approach by \hyperlink{Birrer:2015}{Birrer et a. (2015)}.
The user and developer base of \textsc{lenstronomy} has substantially grown since then, and the software has become an integral part of a wide range of recent analyses, such as measuring the Hubble constant with time-delay strong lensing or constraining the nature of dark matter from resolved and unresolved small scale lensing distortion statistics. 
The modular design has allowed the community to incorporate innovative new methods, as well as to develop enhanced software and wrappers with more specific aims on top of the \textsc{lenstronomy} API.
Through community engagement and involvement, \textsc{lenstronomy} has become a foundation of an ecosystem of affiliated packages extending the original scope of the software and proving its robustness and applicability
at the forefront of the strong gravitational lensing community in an open source and reproducible manner.

\begin{figure}
\centering
\includegraphics{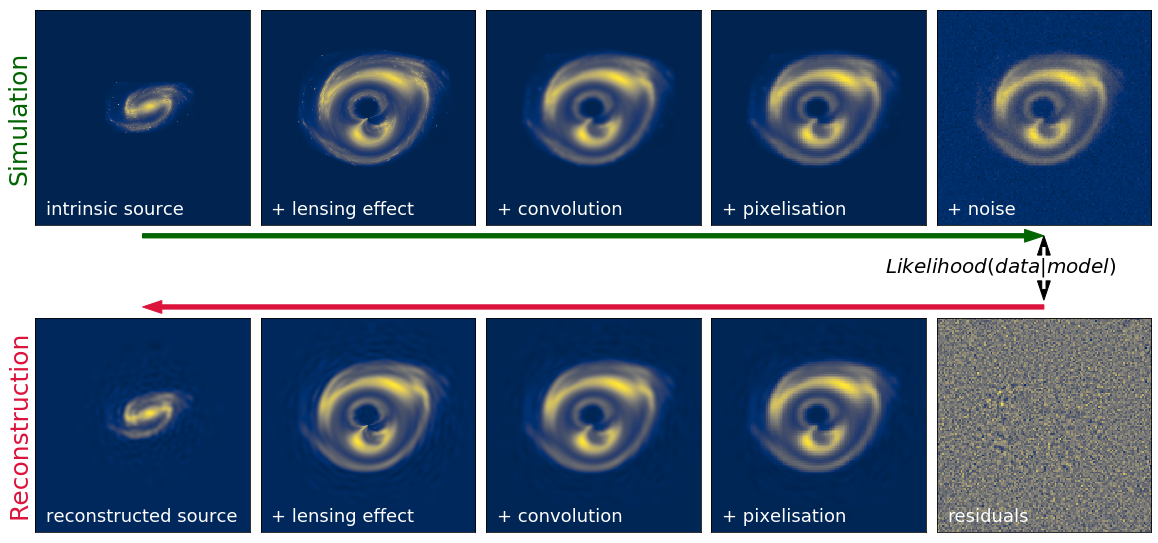}
\caption{Illustration of the strong gravitational lensing phenomenology and the capabilities of \textsc{lenstronomy} in performing realistic simulations as well as reconstructing lensing and source properties from a given data set. Top row from left to right along the green arrow:
A galaxy is lensed around a foreground massive object, becomes highly distorted, and has components appearing multiple times. Observations of this phenomena are limited in resolution (convolution), depending on the detector (pixelation), and are subject to noise.
Bottom row from right to left along the red arrow: The inverse problem is solved with a linear basis set in the source morphology maximizing the likelihood of the model given the data.}
\label{fig:example}
\end{figure}

\section{Background}

Gravitational lensing displaces the observed positions and distorts the shapes of apparent objects on the sky due to intervening inhomogeneous matter along the line of sight. Strong gravitational lensing describes the regime where the background source, such as a galaxy or quasar, is lensed by a massive foreground object, such as another galaxy or cluster of galaxies, to produce multiple images of the source in a highly distorted manner. 
The top row of Figure~\ref{fig:example} illustrates such a process from the intrinsic galaxy to the data product at hand, including the lensing distortions, effects of the instrument, observational conditions, and noise.

Analyses of strong gravitational lensing have provided a wealth of key insights into cosmology and astrophysics.
For example, relative time delays of multiply imaged variable sources provided precision measurements on the expansion rate of the Universe (\hyperlink{Wong:2020}{Wong et al., 2020}; \hyperlink{Shajib:2020strides}{Shajib et al., 2020}; \hyperlink{Birrer:2020tdcosmoiv}{Birrer et al., 2020}). Small scale distortions in the lensing signal of resolved sources 
(\hyperlink{Vegetti:2012}{Vegetti et al., 2012}; \hyperlink{Hezaveh:2016}{Hezaveh et al., 2016}; \hyperlink{Birrer:2017}{Birrer et al., 2017})
and unresolved flux ratios
(\hyperlink{Gilman:2020}{Gilman et al. 2020}; \hyperlink{Hsueh:2020}{Hsueh et al. 2020}) constrain the nature of dark matter. Combined strong lensing and kinematic observables constrain the formation and evolution of galaxies
(\hyperlink{Sonnenfeld:2015}{Sonnenfeld et al., 2015}; \hyperlink{Shajib:2021slacs}{Shajib et al. 2021a})
, and the lensing magnification effect provides an otherwise inaccessible angle on the early Universe (\hyperlink{Zheng:2012}{Zheng et al. 2012}; \hyperlink{Cava:2018}{Cava et al. 2018}).

\section{Statement of need}\label{statement-of-need}

Strong lensing studies have significantly enhanced, and sometimes challenged, our current fundamental understanding of the Universe.
In the near future, with the onset of the next-generation ground and space-based wide and deep astronomical imaging 
(Rubin, Roman, Euclid observatories; \hyperlink{LSST}{Ivezić et al., 2019}; \hyperlink{Roman}{Spergel et al., 2013}; \hyperlink{Euclid}{Laureijs et al., 2011} and interferometric
(SKA; \hyperlink{SKA}{Dewdney et al., 2009}) surveys, the number of discovered lenses of different types will be growing by more than an order of magnitude (\hyperlink{Collett:2015}{Collett 2015}; \hyperlink{OM10}{Oguri \& Marshall, 2010}).
Such large samples can provide unprecedented statistical precision to stress-test our current understanding and exploit discovery potential.
It is key that these demanding studies, at present and in the future, are conducted by reliable software and supported by reproducible and open-source analysis products to provide the most compelling and transparent evidence required to further our physical understanding.

The primary design goal of \textsc{lenstronomy} is to facilitate scientific investigations into the outstanding and most pressing questions in the cosmology and astrophysics community.
\textsc{lenstronomy} has been applied throughout its development to the most demanding modeling and inference problems in strong lensing and the software has evolved around the requirements of the scientific applications to facilitate robust analyses. The modular API of the original design of \textsc{lenstronomy} (\hyperlink{lenstronomy1}{Birrer \& Amara 2018}) has accommodated the addition of new features. Code review processes in the development phase have led to additional benefits for the user community at large beyond the specific needs of the developer.

\textsc{lenstronomy} provides reliable and well-tested specific functionalities, as well as top-level interfaces, which allow for adaptive and innovative usage in control by the scientific investigator.
Guidance for the user community is provided on multiple levels. First, source code is well documented and provided through
\href{http://lenstronomy.readthedocs.org}{readthedocs.org}. Second, a set of \textsc{jupyter} notebooks are provided in an \href{https://github.com/sibirrer/lenstronomy_extensions}{extension repository}. These notebooks demonstrate simplified example use cases, each notebook individually highlighting different specific functionalities of \textsc{lenstronomy}, including a \href{https://github.com/sibirrer/lenstronomy_extensions/blob/v1.8.1/lenstronomy_extensions/Notebooks/starting_guide.ipynb}{starting guide notebook}  to introduce the modular design structure of \textsc{lenstronomy}. Third, end-to-end analysis pipelines of some of the published work are publicly available, providing ‘real-life’ examples at advanced levels.

\section{Track-record of applications}

\textsc{lenstronomy} has been applied in and contributed to more than 30 peer reviewed publications since its first public release in 2018.
In particular, \textsc{lenstronomy} has been used to provide state-of-the-art measurements on real data sets, such as: 
(i) Hubble constant measurements from three quadruly lensed quasars with Hubble Space Telescope (HST) imaging (\hyperlink{Birrer:2016}{Birrer et al., 2016}; \hyperlink{Birrer:2019}{Birrer et al., 2019}; \hyperlink{Shajib:2020strides}{Shajib et al., 2020}), 
dynamical modeling in the hierarchical analysis by \hyperlink{Birrer:2020tdcosmoiv}{Birrer et al., (2020)}, and modeling of lensed supernovae (\hyperlink{Moertsell:2020}{Mörtsell et al. 2020}); 
(ii) inference of small scale dark matter properties from detailed studies of both, resolved imaging (\hyperlink{Birrer:2017}{Birrer et al., 2017}), and unresolved flux ratio statistics (\hyperlink{Gilman:2020}{Gilman et al., 2020}); 
(iii) decomposition of quasar and host galaxy light in both, lensed and unlensed cases (\hyperlink{Ding:2020}{Ding et al., 2020}; \hyperlink{Bennert:2021}{Bennert et al., 2021}); 
(iv) morphological studies of high-redshift sources in the cluster environment
(\hyperlink{Yang:2020}{Yang et al., 2020}; \hyperlink{Yang:2021}{2021}); 
(v) internal structure of galaxies (\hyperlink{Shajib:2021slacs}{Shajib et al., 2021a}; \hyperlink{Shajib:2021AO}{2021b}); 
(vi) measurements of the weak lensing effect imprinted in Einstein rings
(\hyperlink{Birrer:2017cosmos}{Birrer et al., 2017}; \hyperlink{Kuhn:2021}{Kuhn et al., 2021}).
Among the studies, some of them have applied a pipeline to uniformly analyse dozens of lenses of different types
(\hyperlink{Shajib:2019strides}{Shajib et al., 2019}; \hyperlink{Shajib:2021slacs}{2021a}; \hyperlink{Shajib:2021AO}{2021b}), 
a milestone in moving towards utilizing thousands of lenses in the near future.

Beyond analyzing data, many theoretical studies have been conducted using \textsc{lenstronomy} to investigate statistical robustness in present and anticipated future analyses 
(\hyperlink{BirrerTreu:2019}{Birrer \& Treu, 2019}; \hyperlink{Millon:2020}{Millon et al., 2020}; \hyperlink{vdVyvere:2020}{Van de Vyvere et al., 2020}; \hyperlink{Li:2021}{Li et al., 2021}; \hyperlink{Ding:2021transient}{Ding et al. 2021}), 
as well as to provide forecasts for anticipated future constraints for different science cases (\hyperlink{Gilman:2019}{Gilman et al., 2019}; \hyperlink{Sengul:2020}{Çaǧan Şengül et al. 2020}; \hyperlink{BirrerTreu:2021}{Birrer \& Treu, 2021}).
Particularly, three separate teams participated in the blind time-delay lens modeling challenge (\hyperlink{Ding:2021tdlmc}{Ding et al. 2021}) using \textsc{lenstronomy}.

\textsc{lenstronomy} has seen a substantial development and incorporation of innovations and numerical recipes 
(\hyperlink{Tessore:2015}{Tessore \& Metcalf, 2015}; \hyperlink{Shajib:2019unified}{Shajib, 2019}; \hyperlink{Joseph:2019}{Joseph et al., 2019}; \hyperlink{Galan:2021}{Galan et al., 2021}; \hyperlink{Birrer:2021arcs}{Birrer, 2021}), 
and has found applications beyond its original aim due to the robust and high-standard design requirements.

\section{Ecosystem of affiliated packages}

\textsc{lenstronomy} has allowed the community to develop third-party analysis products and software products utilizing its core functionalities to provide more targeted and integrated software solutions for a wide range of scientific analyses. 
These open-source \href{https://github.com/sibirrer/lenstronomy/blob/1.8.1/AFFILIATEDPACKAGES.rst}{affiliated packages} effectively create an ecosystem enhancing the capability of \textsc{lenstronomy}. 
They provide specified and tested solution for specific scientific investigations, such as plug-ins and direct implementation for innovative source reconstruction algorithms (\href{https://github.com/aymgal/SLITronomy}{\textsc{SLITronomy}}; \hyperlink{Joseph:2019}{Joseph et al., 2019}; \hyperlink{Galan:2021}{Galan et al., 2021}), 
gravitational wave lensing computations (\href{https://gitlab.com/gpagano/lensinggw}{\textsc{lensingGW}}; \hyperlink{Pagano:2020}{Pagano et al., 2020}), 
automated pipelines for gravitational lensing reconstruction
(\href{https://github.com/ajshajib/dolphin}{\textsc{dolphin}}; \hyperlink{Shajib:2021slacs}{Shajib et al., 2021a}), 
cluster source reconstruction and local perturbative lens modeling
(\href{https://github.com/ylilan/lenstruction}{\textsc{lenstruction}}; \hyperlink{Yang:2020}{Yang et al., 2020}), 
enhancement in large-scale structure imaging survey simulations
(\href{https://github.com/LSSTDESC/SLSprinkler}{DESC \textsc{SLSprinkler}}; \hyperlink{LSSTDESC:2021}{Dark Energy Science Collaboration (LSST DESC) et al., 2021}), 
rendering of sub-halos and line-of-sight halos
(\href{https://github.com/dangilman/pyHalo}{\textsc{pyHalo}}; \hyperlink{Gilman:2020}{Gilman et al., 2020}),
galaxy morphology analysis (\href{https://github.com/dartoon/galight}{\textsc{galight}}; \hyperlink{Ding:2020}{Ding et al., 2020}),
and hierarchical analyses to measure the Hubble constant
(\href{https://github.com/sibirrer/hierarc}{\textsc{hierArc}}; \hyperlink{Birrer:2020tdcosmoiv}{Birrer et al., 2020}).
With the rise in popularity and the promises in dealing with ever complex data problems with fast deep-learning methods, 
dedicated tools for simulating large datasets for applying such methods to strong gravitational lensing
(\href{https://github.com/deepskies/deeplenstronomy}{\textsc{deeplenstronomy}}; \hyperlink{Morgan:2021}{Morgan et al., 2021}),
(\href{https://github.com/jiwoncpark/baobab}{\textsc{baobab}}; \hyperlink{Park:2021}{Park et al., 2021}), 
as well as end-to-end Bayesian Neural Network training and validation packages for Hubble constant measurements
(\href{https://github.com/jiwoncpark/h0rton}{\textsc{h0rton}}; \hyperlink{Park:2021}{Park et al., 2021}), 
and for a hierarchical analysis of galaxy-galaxy lenses
(\href{https://github.com/swagnercarena/ovejero}{\textsc{ovejero}}; \hyperlink{Wagner-Carena:2021}{Wagner-Carena et al., 2021}) have been developed.
The affiliated packages make best use of the \textsc{lenstronomy} modules without duplicating source code and make it possible to combine aspects of multiple affiliated packages in one single analysis.

\section{Related open source software}
\begin{itemize}
    \item \href{https://github.com/sibirrer/lenstronomy}{\textsc{lenstronomy}}(\hyperlink{Birrer:2015}{Birrer et al., 2015}; \hyperlink{lenstronomy1}{Birrer \& Amara 2018})
    \item \href{https://github.com/Jammy2211/PyAutoLens}{\textsc{PyAutoLens}}(\hyperlink{Nightingale:2018}{Nightingale et al., 2018}; \hyperlink{Nightingale:2021}{2021})
    \item \href{http://www.physics.rutgers.edu/~keeton/gravlens/}{\textsc{gravlens}}(\hyperlink{Keeton:2011}{Keeton, 2011})
    \item \href{https://www.slac.stanford.edu/~oguri/glafic/}{\textsc{glafic}}(\hyperlink{Oguri:2010}{Oguri, 2010})
    \item \href{https://github.com/jspilker/visilens}{\textsc{visilens}} (\hyperlink{Spilker:2016}{Spilker et al., 2016})
    \item \href{https://www.physik.uzh.ch/~psaha/lens/pixelens.php}{\textsc{PixeLens}} (\hyperlink{PixeLens}{Saha \& Williams, 2011})
    \item \href{https://github.com/j0r1/GRALE2}{\textsc{grale}} (\hyperlink{GRALE}{Liesenborgs et al., 2006})
    \item \href{http://projets.lam.fr/projects/lenstool/wiki}{\textsc{lenstool}} (\hyperlink{Jullo:2009}{Jullo \& Kneib, 2009})
\end{itemize}

\section{Acknowledgements}\label{acknowledgements}

Support for this work was provided by the National Science Foundation through NSF AST-1716527. 
AJS was supported by NASA through the STScI grant HST-GO-15320 and by a Dissertation Year Fellowship from the UCLA Graduate Division. 
This research was supported by the U.S. Department of Energy (DOE) Office of Science Distinguished Scientist Fellow Program.
DG is supported by NASA HST-GO-15177.
AG, MM LvdV, DS are supported by COSMICLENS: ERC grant agreement No 787886.
LT is supported by International Helmholtz-Weizmann Research School for Multimessenger Astronomy.
MU is supported by KIPAC and the Stanford Summer Research Program.
XD is supported by NASA HST-GO-15115.
TS is supported by NASA grant HST-GO-15320 and HST-GO-15652.
MZ is supported by the National Science Foundation of China.
AA is supported by a Royal Society Wolfson Fellowship.
We are grateful to the user community for valuable feedback and encouragement in continuing the development.

\section*{References}\label{references}
\addcontentsline{toc}{section}{References}

\hypertarget{refs}{}

\hypertarget{astropy:2013}{Astropy Collaboration, Robitaille, T. P., Tollerud, E. J., Greenfield, P., Droettboom, M., Bray,
E., Aldcroft, T., Davis, M., Ginsburg, A., Price-Whelan, A. M., Kerzendorf, W. E., Conley,
A., Crighton, N., Barbary, K., Muna, D., Ferguson, H., Grollier, F., Parikh, M. M., Nair,
P. H., … Streicher, O. (2013). Astropy: A community Python package for astronomy.
\textit{Astronomy \& Astrophysics, 558, A33}. \href{https://doi.org/10.1051/0004-6361/201322068}{https://doi.org/10.1051/0004-6361/201322068}}

\hypertarget{astropy:2018}{
Astropy Collaboration, Price-Whelan, A. M., Sipőcz, B. M., Günther, H. M., Lim, P. L.,
Crawford, S. M., Conseil, S., Shupe, D. L., Craig, M. W., Dencheva, N., Ginsburg, A.,
Vand erPlas, J. T., Bradley, L. D., Pérez-Suárez, D., de Val-Borro, M., Aldcroft, T. L.,
Cruz, K. L., Robitaille, T. P., Tollerud, E. J., … Astropy Contributors. (2018). The
Astropy Project: Building an Open-science Project and Status of the v2.0 Core Package.
\textit{The Astronomical Journal, 156(3), 123}. \href{https://doi.org/10.3847/1538-3881/aabc4f}{https://doi.org/10.3847/1538-3881/aabc4f}}

\hypertarget{Bennert:2021}{
Bennert, V. N., Treu, T., Ding, X., Stomberg, I., Birrer, S., Snyder, T., Malkan, M. A.,
Stephens, A. W., \& Auger, M. W. (2021). A local baseline of the black hole mass scaling relations for active galaxies. IV. Correlations between $M_{\rm BH}$ and host galaxy $\sigma$, stellar
mass, and luminosity. \textit{arXiv e-Prints}, arXiv:2101.10355. \href{http://arxiv.org/abs/2101.10355}{http://arxiv.org/abs/2101.10355}}

\hypertarget{Birrer:2021arcs}{Birrer, S. (2021). Gravitational lensing formalism in a curved arc basis: A continuous description
of observables and degeneracies from the weak to the strong lensing regime. \textit{arXiv e-Prints}, arXiv:2104.09522. \href{http://arxiv.org/abs/2104.09522}{http://arxiv.org/abs/2104.09522}}

\hypertarget{lenstronomy1}{Birrer, S., \& Amara, A. (2018). lenstronomy: Multi-purpose gravitational lens modelling
software package. \textit{Physics of the Dark Universe, 22, 189–201}. \href{https://doi.org/10.1016/j.dark.2018.11.002}{https://doi.org/10.1016/j.dark.2018.11.002}}

\hypertarget{Birrer:2017}{
Birrer, S., Amara, A., \& Refregier, A. (2017). Lensing substructure quantification in RXJ1131-
1231: a 2 keV lower bound on dark matter thermal relic mass. \textit{Journal of Cosmology and
Astroparticle Physics, 2017(5), 037}. \href{https://doi.org/10.1088/1475-7516/2017/05/037}{https://doi.org/10.1088/1475-7516/2017/05/037}}

\hypertarget{Birrer:2016}{
Birrer, S., Amara, A., \& Refregier, A. (2016). The mass-sheet degeneracy and time-delay
cosmography: analysis of the strong lens RXJ1131-1231. 
\textit{Journal of Cosmology and
Astroparticle Physics, 2016(8), 020}. \href{https://doi.org/10.1088/1475-7516/2016/08/020}{https://doi.org/10.1088/1475-7516/2016/08/020}}

\hypertarget{Birrer:2015}{
Birrer, S., Amara, A., \& Refregier, A. (2015). Gravitational Lens Modeling with Basis Sets.
\textit{The Astrophysical Journal, 813(2), 102}. \href{https://doi.org/10.1088/0004-637X/813/2/102}{https://doi.org/10.1088/0004-637X/813/2/102}}

\hypertarget{Birrer:2020tdcosmoiv}{
Birrer, S., Shajib, A. J., Galan, A., Millon, M., Treu, T., Agnello, A., Auger, M., Chen, G.
C.-F., Christensen, L., Collett, T., Courbin, F., Fassnacht, C. D., Koopmans, L. V. E.,
Marshall, P. J., Park, J.-W., Rusu, C. E., Sluse, D., Spiniello, C., Suyu, S. H., … Van de
Vyvere, L. (2020). TDCOSMO. IV. Hierarchical time-delay cosmography joint inference
of the Hubble constant and galaxy density profiles. 
\textit{Astronomy \& Astrophysics, 643, A165}.
\href{https://doi.org/10.1051/0004-6361/202038861}{https://doi.org/10.1051/0004-6361/202038861}}

\hypertarget{BirrerTreu:2021}{
Birrer, S., \& Treu, T. (2021). TDCOSMO V: strategies for precise and accurate measurements of the Hubble constant with strong lensing. 
\textit{Astronomy \& Astrophysics, 649, A61, 6}.
\href{https://doi.org/10.1051/0004-6361/202039179}{https://doi.org/10.1051/0004-6361/202039179}}

\hypertarget{BirrerTreu:2019}{
Birrer, S., \& Treu, T. (2019). Astrometric requirements for strong lensing time-delay cosmography.
\textit{Monthly Notices of the Royal Astronomical Society, 489(2), 2097–2103}.
\href{https://doi.org/10.1093/mnras/stz2254}{https://doi.org/10.1093/mnras/stz2254}}

\hypertarget{Birrer:2019}{
Birrer, S., Treu, T., Rusu, C. E., Bonvin, V., Fassnacht, C. D., Chan, J. H. H., Agnello,
A., Shajib, A. J., Chen, G. C.-F., Auger, M., Courbin, F., Hilbert, S., Sluse, D., Suyu,
S. H., Wong, K. C., Marshall, P., Lemaux, B. C., \& Meylan, G. (2019). H0LiCOW
- IX. Cosmographic analysis of the doubly imaged quasar SDSS 1206+4332 and a new measurement of the Hubble constant. 
\textit{Monthly Notices of the Royal Astronomical Society,
484(4), 4726–4753}. \href{https://doi.org/10.1093/mnras/stz200}{https://doi.org/10.1093/mnras/stz200}}

\hypertarget{Birrer:2017cosmos}{
Birrer, S., Welschen, C., Amara, A., \& Refregier, A. (2017). Line-of-sight effects in strong lensing: putting theory into practice. 
\textit{Journal of Cosmology and Astroparticle Physics,
2017(4), 049}. \href{https://doi.org/10.1088/1475-7516/2017/04/049}{https://doi.org/10.1088/1475-7516/2017/04/049}}

\hypertarget{Cava:2018}{
Cava, A., Schaerer, D., Richard, J., Pérez-González, P. G., Dessauges-Zavadsky, M., Mayer,
L., \& Tamburello, V. (2018). The nature of giant clumps in distant galaxies probed by the anatomy of the cosmic snake. 
\textit{Nature Astronomy, 2, 76–82}. \href{https://doi.org/10.1038/
s41550-017-0295-x}{https://doi.org/10.1038/
s41550-017-0295-x}}

\hypertarget{Collett:2015}{
Collett, T. E. (2015). The Population of Galaxy-Galaxy Strong Lenses in Forthcoming Optical Imaging Surveys. 
\textit{The Astrophysical Journal, 811(1), 20}. \href{https://doi.org/10.1088/0004-637X/811/1/20}{https://doi.org/10.1088/0004-637X/811/1/20}}

\hypertarget{Sengul:2020}{
Çaǧan Şengül, A., Tsang, A., Diaz Rivero, A., Dvorkin, C., Zhu, H.-M., \& Seljak, U. (2020). Quantifying the line-of-sight halo contribution to the dark matter convergence power spectrum from strong gravitational lenses. \textit{Physical Review D, 102(6), 063502}.
\href{https://doi.org/10.1103/PhysRevD.102.063502}{https://doi.org/10.1103/PhysRevD.102.063502}}

\hypertarget{SKA}{
Dewdney, P. E., Hall, P. J., Schilizzi, R. T., 
\& Lazio, T. J. L. W. (2009). The Square
Kilometre Array. 
\textit{IEEE Proceedings, 97(8), 1482–1496}. \href{https://doi.org/10.1109/JPROC.2009.2021005}{https://doi.org/10.1109/JPROC.2009.2021005}}

\hypertarget{Ding:2021transient}{
Ding, X., Liao, K., Birrer, S., Shajib, A. J., Treu, T., \& Yang, L. (2021). Improved time delay lens modelling and H0 inference with transient sources. \textit{Monthly Notices of the Royal Astronomical Society 504(4), 5621–5628}. \href{https://doi.org/10.1093/mnras/stab1240}{https://doi.org/10.1093/mnras/stab1240}}

\hypertarget{Ding:2020}{
Ding, X., Silverman, J., Treu, T., Schulze, A., Schramm, M., Birrer, S., Park, D., Jahnke,
K., Bennert, V. N., Kartaltepe, J. S., Koekemoer, A. M., Malkan, M. A., \& Sanders, D. (2020). The Mass Relations between Supermassive Black Holes and Their Host Galaxies at 1 < z < 2 HST-WFC3. 
\textit{The Astrophysical Journal, 888(1), 37}. \href{https://doi.org/10.3847/1538-4357/ab5b90}{https://doi.org/10.3847/1538-4357/ab5b90}}

\hypertarget{Ding:2021tdlmc}{
Ding, X., Treu, T., Birrer, S., Chen, G. C.-F., Coles, J., Denzel, P., Frigo, M., Galan, A.,
Marshall, P. J., Millon, M., More, A., Shajib, A. J., Sluse, D., Tak, H., Xu, D., Auger,
M. W., Bonvin, V., Chand, H., Courbin, F., … Williams, L. L. R. (2021). Time delay lens modelling challenge. 
\textit{Monthly Notices of the Royal Astronomical Society, 503(1),
1096–1123}. 
\href{https://doi.org/10.1093/mnras/stab484}{https://doi.org/10.1093/mnras/stab484}}

\hypertarget{Galan:2021}{
Galan, A., Peel, A., Joseph, R., Courbin, F., \& Starck, J.-L. (2021). SLITRONOMY: Towards a fully wavelet-based strong lensing inversion technique. 
\textit{Astronomy \& Astrophysics, 647, A176}. \href{https://doi.org/10.1051/0004-6361/202039363}{https://doi.org/10.1051/0004-6361/202039363}}

\hypertarget{Gilman:2020}{
Gilman, D., Birrer, S., Nierenberg, A., Treu, T., Du, X., \& Benson, A. (2020). Warm dark matter chills out: constraints on the halo mass function and the free-streaming length of dark matter with eight quadruple-image strong gravitational lenses. 
\textit{Monthly Notices of the
Royal Astronomical Society, 491(4), 6077–6101}. \href{https://doi.org/10.1093/mnras/stz3480}{https://doi.org/10.1093/mnras/stz3480}}

\hypertarget{Gilman:2019}{
Gilman, D., Birrer, S., Treu, T., Nierenberg, A., \& Benson, A. (2019). Probing dark matter
structure down to 107 solar masses: flux ratio statistics in gravitational lenses with line-of-sight haloes. 
\textit{Monthly Notices of the Royal Astronomical Society, 487(4), 5721–5738}.
\href{https://doi.org/10.1093/mnras/stz1593}{https://doi.org/10.1093/mnras/stz1593}}

\hypertarget{Hezaveh:2016}{
Hezaveh, Y. D., Dalal, N., Marrone, D. P., Mao, Y.-Y., Morningstar, W., Wen, D., Blandford,
R. D., Carlstrom, J. E., Fassnacht, C. D., Holder, G. P., Kemball, A., Marshall, P. J., Murray, N., Perreault Levasseur, L., Vieira, J. D., \& Wechsler, R. H. (2016). Detection of Lensing Substructure Using ALMA Observations of the Dusty Galaxy SDP.81. 
\textit{The Astrophysical Journal, 823(1), 37}. \href{https://doi.org/10.3847/0004-637X/823/1/37}{https://doi.org/10.3847/0004-637X/823/1/37}}

\hypertarget{Hsueh:2020}{
Hsueh, J.-W., Enzi, W., Vegetti, S., Auger, M. W., Fassnacht, C. D., Despali, G., Koopmans,
L. V. E., \& McKean, J. P. (2020). SHARP - VII. New constraints on the dark matter free-streaming properties and substructure abundance from gravitationally lensed quasars.
\textit{Monthly Notices of the Royal Astronomical Society, 492(2), 3047–3059}. 
\href{https://doi.org/10.1093/mnras/stz3177}{https://doi.org/10.1093/mnras/stz3177}}

\hypertarget{LSST}{
Ivezić, Ž., Kahn, S. M., Tyson, J. A., Abel, B., Acosta, E., Allsman, R., Alonso, D., AlSayyad,
Y., Anderson, S. F., Andrew, J., Angel, J. R. P., Angeli, G. Z., Ansari, R., Antilogus,
P., Araujo, C., Armstrong, R., Arndt, K. T., Astier, P., Aubourg, É., … Zhan, H. (2019).
LSST: From Science Drivers to Reference Design and Anticipated Data Products. 
\textit{The Astrophysical Journal, 873(2), 111}. \href{https://doi.org/10.3847/1538-4357/ab042c}{https://doi.org/10.3847/1538-4357/ab042c}}

\hypertarget{Joseph:2019}{
Joseph, R., Courbin, F., Starck, J.-L., \& Birrer, S. (2019). Sparse Lens Inversion Technique (SLIT): lens and source separability from linear inversion of the source reconstruction problem. 
\textit{Astronomy \& Astrophysics, 623, A14}. \href{https://doi.org/10.1051/0004-6361/201731042}{https://doi.org/10.1051/0004-6361/201731042}}

\hypertarget{Jullo:2009}{
Jullo, E., \& Kneib, J.-P. (2009). Multiscale cluster lens mass mapping - I. Strong lensing modelling. 
\textit{Monthly Notices of the Royal Astronomical Society 395(3), 1319–1332}. \href{https://doi.org/10.1111/j.1365-2966.2009.14654.x}{https://doi.org/10.1111/j.1365-2966.2009.14654.x}}

\hypertarget{Keeton:2011}{
Keeton, C. R. (2011). GRAVLENS: Computational Methods for Gravitational Lensing (ascl:1102.003).}

\hypertarget{Kuhn:2021}{
Kuhn, F. A., Birrer, S., Bruderer, C., Amara, A., \& Refregier, A. (2021). Combining strong and weak lensing estimates in the Cosmos field. 
\textit{Journal of Cosmology and Astroparticle
Physics, 2021(4), 010}. \href{https://doi.org/10.1088/1475-7516/2021/04/010}{https://doi.org/10.1088/1475-7516/2021/04/010}}

\hypertarget{Euclid}{
Laureijs, R., Amiaux, J., Arduini, S., Auguères, J.-L., Brinchmann, J., Cole, R., Cropper, M.,
Dabin, C., Duvet, L., Ealet, A., Garilli, B., Gondoin, P., Guzzo, L., Hoar, J., Hoekstra, H.,
Holmes, R., Kitching, T., Maciaszek, T., Mellier, Y., … Zucca, E. (2011). Euclid Definition Study Report. 
\textit{arXiv e-Prints}, arXiv:1110.3193. 
\href{http://arxiv.org/abs/1110.3193}{http://arxiv.org/abs/1110.3193}}

\hypertarget{Li:2021}{
Li, N., Becker, C., \& Dye, S. (2021). The impact of line-of-sight structures on measuring H0 with strong lensing time-delays. 
\textit{Monthly Notices of the Royal Astronomical Society}.
\href{https://doi.org/10.1093/mnras/stab984}{https://doi.org/10.1093/mnras/stab984}}

\hypertarget{GRALE}{
Liesenborgs, J., De Rijcke, S., \& Dejonghe, H. (2006). A genetic algorithm for the nonparametric inversion of strong lensing systems. 
\textit{Monthly Notices of the Royal Astronomical Society, 367(3), 1209–1216}. 
\href{https://doi.org/10.1111/j.1365-2966.2006.10040.x}{https://doi.org/10.1111/j.1365-2966.2006.10040.x}}

\hypertarget{LSSTDESC:2021}{
LSST Dark Energy Science Collaboration (LSST DESC), Abolfathi, B., Alonso, D., Armstrong,
R., Aubourg, É., Awan, H., Babuji, Y. N., Bauer, F. E., Bean, R., Beckett, G., Biswas,
R., Bogart, J. R., Boutigny, D., Chard, K., Chiang, J., Claver, C. F., Cohen-Tanugi, J.,
Combet, C., Connolly, A. J., … Zuntz, J. (2021). The LSST DESC DC2 Simulated Sky Survey. 
\textit{The Astrophysical Journal Supplement Series, 253(1), 31}. \href{https://doi.org/10.3847/1538-4365/abd62c}{https://doi.org/10.3847/1538-4365/abd62c}}

\hypertarget{Millon:2020}{
Millon, M., Galan, A., Courbin, F., Treu, T., Suyu, S. H., Ding, X., Birrer, S., Chen, G.
C.-F., Shajib, A. J., Sluse, D., Wong, K. C., Agnello, A., Auger, M. W., Buckley-Geer,
E. J., Chan, J. H. H., Collett, T., Fassnacht, C. D., Hilbert, S., Koopmans, L. V. E., …
Van de Vyvere, L. (2020). TDCOSMO. I. An exploration of systematic uncertainties in the inference of H0 from time-delay cosmography. 
\textit{Astronomy \& Astrophysics, 639, A101}.
\href{https://doi.org/10.1051/0004-6361/201937351}{https://doi.org/10.1051/0004-6361/201937351}}

\hypertarget{Morgan:2021}{
Morgan, R., Nord, B., Birrer, S., Lin, J., \& Poh, J. (2021). deeplenstronomy: A dataset simulation package for strong gravitational lensing. 
\textit{The Journal of Open Source Software,
6(58), 2854}. \href{https://doi.org/10.21105/joss.02854}{https://doi.org/10.21105/joss.02854}}

\hypertarget{Moertsell:2020}{
Mörtsell, E., Johansson, J., Dhawan, S., Goobar, A., Amanullah, R., \& Goldstein, D. A.
(2020). Lens modelling of the strongly lensed Type Ia supernova iPTF16geu. \textit{Monthly Notices of the Royal Astronomical Society, 496(3), 3270–3280}. \href{https://doi.org/10.1093/mnras/staa1600}{https://doi.org/10.1093/mnras/staa1600}}

\hypertarget{Nightingale:2018}{
Nightingale, J. W., Dye, S., \& Massey, R. J. (2018). AutoLens: automated modeling of a strong lens’s light, mass, and source. 
\textit{Monthly Notices of the Royal Astronomical Society, 478(4), 4738–4784}. \href{https://doi.org/10.1093/mnras/sty1264}{https://doi.org/10.1093/mnras/sty1264}}

\hypertarget{Nightingale:2021}{
Nightingale, J. W., Hayes, R., Kelly, A., Amvrosiadis, A., Etherington, A., He, Q., Li, N., Cao,
X., Frawley, J., Cole, S., Enia, A., Frenk, C., Harvey, D., Li, R., Massey, R., Negrello, M., \& Robertson, A. (2021). PyAutoLens: Open-Source Strong Gravitational Lensing. 
\textit{The Journal of Open Source Software, 6(58), 2825}. \href{https://doi.org/10.21105/joss.02825}{https://doi.org/10.21105/joss.02825}}

\hypertarget{Oguri:2010}{
Oguri, M. (2010). glafic: Software Package for Analyzing Gravitational Lensing (ascl:1010.012).}

\hypertarget{OM10}{
Oguri, M., \& Marshall, P. J. (2010). Gravitationally lensed quasars and supernovae in future wide-field optical imaging surveys. 
\textit{Monthly Notices of the Royal Astronomical Society,
405(4), 2579–2593}. \href{https://doi.org/10.1111/j.1365-2966.2010.16639.x}{https://doi.org/10.1111/j.1365-2966.2010.16639.x}}

\hypertarget{Pagano:2020}{
Pagano, G., Hannuksela, O. A., \& Li, T. G. F. (2020). LENSINGGW: a PYTHON package for lensing of gravitational waves. 
\textit{Astronomy \& Astrophysics, 643, A167}. 
\href{https://doi.org/10.1051/0004-6361/202038730}{https://doi.org/10.1051/0004-6361/202038730}}

\hypertarget{Park:2021}{
Park, J. W., Wagner-Carena, S., Birrer, S., Marshall, P. J., Lin, J. Y.-Y., Roodman, A., \&
LSST Dark Energy Science Collaboration. (2021). Large-scale Gravitational Lens Modeling with Bayesian Neural Networks for Accurate and Precise Inference of the Hubble Constant. 
\textit{The Astrophysical Journal, 910(1), 39}. \href{https://doi.org/10.3847/1538-4357/abdfc4}{https://doi.org/10.3847/1538-4357/abdfc4}}

\hypertarget{PixeLens}{
Saha, P., \& Williams, L. L. R. (2011). PixeLens: A Portable Modeler of Lensed Quasars (ascl:1102.007).}

\hypertarget{Shajib:2019unified}{
Shajib, A. J. (2019). Unified lensing and kinematic analysis for any elliptical mass profile.
\textit{Monthly Notices of the Royal Astronomical Society, 488(1), 1387–1400}. 
\href{https://doi.org/10.1093/mnras/stz1796}{https://doi.org/10.1093/mnras/stz1796}}

\hypertarget{Shajib:2020strides}{
Shajib, A. J., Birrer, S., Treu, T., Agnello, A., Buckley-Geer, E. J., Chan, J. H. H., Christensen,
L., Lemon, C., Lin, H., Millon, M., Poh, J., Rusu, C. E., Sluse, D., Spiniello, C., Chen,
G. C.-F., Collett, T., Courbin, F., Fassnacht, C. D., Frieman, J., … Zhang, Y. (2020).
STRIDES: a 3.9 per cent measurement of the Hubble constant from the strong lens system DES J0408-5354. 
\textit{Monthly Notices of the Royal Astronomical Society, 494(4),
6072–6102}. \href{https://doi.org/10.1093/mnras/staa828}{https://doi.org/10.1093/mnras/staa828}}

\hypertarget{Shajib:2019strides}{
Shajib, A. J., Birrer, S., Treu, T., Auger, M. W., Agnello, A., Anguita, T., Buckley-Geer, E.
J., Chan, J. H. H., Collett, T. E., Courbin, F., Fassnacht, C. D., Frieman, J., Kayo, I.,
Lemon, C., Lin, H., Marshall, P. J., McMahon, R., More, A., Morgan, N. D., … Walker,
A. R. (2019). Is every strong lens model unhappy in its own way? Uniform modelling of a sample of 13 quadruply+ imaged quasars. 
\textit{Monthly Notices of the Royal Astronomical
Society, 483(4), 5649–5671}. 
\href{https://doi.org/10.1093/mnras/sty3397}{https://doi.org/10.1093/mnras/sty3397}}

\hypertarget{Shajib:2021AO}{
Shajib, A. J., Molina, E., Agnello, A., Williams, P. R., Birrer, S., Treu, T., Fassnacht, C.
D., Morishita, T., Abramson, L., Schechter, P. L., \& Wisotzki, L. (2021). High-resolution imaging follow-up of doubly imaged quasars. 
\textit{Monthly Notices of the Royal Astronomical
Society, 503(2), 1557–1567}. \href{https://doi.org/10.1093/mnras/stab532}{https://doi.org/10.1093/mnras/stab532}}

\hypertarget{Shajib:2021slacs}{
Shajib, A. J., Treu, T., Birrer, S., \& Sonnenfeld, A. (2021). Dark matter haloes of massive elliptical galaxies at z $\sim$ 0.2 are well described by the Navarro-Frenk-White profile. 
\textit{Monthly Notices of the Royal Astronomical Society, 503(2), 2380–2405}. \href{https://doi.org/10.1093/mnras/stab536}{https://doi.org/10.1093/mnras/stab536}}

\hypertarget{Sonnenfeld:2015}{
Sonnenfeld, A., Treu, T., Marshall, P. J., Suyu, S. H., Gavazzi, R., Auger, M. W., \& Nipoti,
C. (2015). The SL2S Galaxy-scale Lens Sample. V. Dark Matter Halos and Stellar IMF of Massive Early-type Galaxies Out to Redshift 0.8. 
\textit{The Astrophysical Journal, 800(2),
94}. \href{https://doi.org/10.1088/0004-637X/800/2/94}{https://doi.org/10.1088/0004-637X/800/2/94}}

\hypertarget{Roman}{
Spergel, D., Gehrels, N., Breckinridge, J., Donahue, M., Dressler, A., Gaudi, B. S., Greene,
T., Guyon, O., Hirata, C., Kalirai, J., Kasdin, N. J., Moos, W., Perlmutter, S., Postman,
M., Rauscher, B., Rhodes, J., Wang, Y., Weinberg, D., Centrella, J., … Shaklan, S. (2013).
Wide-Field InfraRed Survey Telescope-Astrophysics Focused Telescope Assets WFIRSTAFTA
Final Report. 
\textit{arXiv e-Prints}, arXiv:1305.5422. 
\href{http://arxiv.org/abs/1305.5422}{http://arxiv.org/abs/1305.5422}}

\hypertarget{Spilker:2016}{
Spilker, J. S., Marrone, D. P., Aravena, M., Béthermin, M., Bothwell, M. S., Carlstrom, J.
E., Chapman, S. C., Crawford, T. M., de Breuck, C., Fassnacht, C. D., Gonzalez, A. H.,
Greve, T. R., Hezaveh, Y., Litke, K., Ma, J., Malkan, M., Rotermund, K. M., Strandet,
M., Vieira, J. D., … Welikala, N. (2016). ALMA Imaging and Gravitational Lens Models of South Pole TelescopeSelected Dusty, Star-Forming Galaxies at High Redshifts. 
\textit{The Astrophysical Journal, 826, 112}. \href{https://doi.org/10.3847/0004-637X/826/2/112}{https://doi.org/10.3847/0004-637X/826/2/112}}

\hypertarget{Tessore:2015}{
Tessore, N., \& Metcalf, R. B. (2015). The elliptical power law profile lens. 
\textit{Astronomy \& Astrophysics, 580, A79}. \href{https://doi.org/10.1051/0004-6361/201526773}{https://doi.org/10.1051/0004-6361/201526773}}

\hypertarget{vdVyvere:2020}{
Van de Vyvere, L., Sluse, D., Mukherjee, S., Xu, D., \& Birrer, S. (2020). The impact of mass
map truncation on strong lensing simulations. 
\textit{Astronomy \& Astrophysics, 644, A108}.
\href{https://doi.org/10.1051/0004-6361/202038942}{https://doi.org/10.1051/0004-6361/202038942}}

\hypertarget{Vegetti:2012}{
Vegetti, S., Lagattuta, D. J., McKean, J. P., Auger, M. W., Fassnacht, C. D., \& Koopmans,
L. V. E. (2012). Gravitational detection of a low-mass dark satellite galaxy at cosmological distance. 
\textit{Nature, 481(7381), 341–343}. 
\href{https://doi.org/10.1038/nature10669}{https://doi.org/10.1038/nature10669}}

\hypertarget{Wagner-Carena:2021}{
Wagner-Carena, S., Park, J. W., Birrer, S., Marshall, P. J., Roodman, A., Wechsler, R. H., \& LSST Dark Energy Science Collaboration. (2021). Hierarchical Inference with Bayesian
Neural Networks: An Application to Strong Gravitational Lensing. 
\textit{The Astrophysical Journal, 909(2), 187}. \href{https://doi.org/10.3847/1538-4357/abdf59}{https://doi.org/10.3847/1538-4357/abdf59}}

\hypertarget{Wong:2020}{
Wong, K. C., Suyu, S. H., Chen, G. C.-F., Rusu, C. E., Millon, M., Sluse, D., Bonvin, V.,
Fassnacht, C. D., Taubenberger, S., Auger, M. W., Birrer, S., Chan, J. H. H., Courbin, F.,
Hilbert, S., Tihhonova, O., Treu, T., Agnello, A., Ding, X., Jee, I., … Meylan, G. (2020).
H0LiCOW XIII. A 2.4 per cent measurement of H0 from lensed quasars: 5.3$\sigma$ tension between early- and late-Universe probes. 
\textit{Monthly Notices of the Royal Astronomical
Society, 498(1), 1420–1439}. \href{https://doi.org/10.1093/mnras/stz3094}{https://doi.org/10.1093/mnras/stz3094}}

\hypertarget{Yang:2020}{
Yang, L., Birrer, S., \& Treu, T. (2020). A versatile tool for cluster lensing source reconstruction - I. Methodology and illustration on sources in the Hubble Frontier Field Cluster MACS
J0717.5+3745. 
\textit{Monthly Notices of the Royal Astronomical Society, 496(3), 2648–2662}.
\href{https://doi.org/10.1093/mnras/staa1649}{https://doi.org/10.1093/mnras/staa1649}}

\hypertarget{Yang:2021}{
Yang, L., Roberts-Borsani, G., Treu, T., Birrer, S., Morishita, T., \& Bradač, M. (2021). The
evolution of the size-mass relation at z = 1-3 derived from the complete Hubble Frontier Fields data set. 
\textit{Monthly Notices of the Royal Astronomical Society, 501(1), 1028–1037}.
\href{https://doi.org/10.1093/mnras/staa3713}{https://doi.org/10.1093/mnras/staa3713}}

\hypertarget{Zheng:2012}{
Zheng, W., Postman, M., Zitrin, A., Moustakas, J., Shu, X., Jouvel, S., H{\o}st, O., Molino,
A., Bradley, L., Coe, D., Moustakas, L. A., Carrasco, M., Ford, H., Ben{\'\i}tez, N., Lauer,
T. R., Seitz, S., Bouwens, R., Koekemoer, A., Medezinski, E., … van der Wel, A. (2012).
A magnified young galaxy from about 500 million years after the Big Bang. \textit{Nature, 489(7416), 406–408}. \href{https://doi.org/10.1038/nature11446}{https://doi.org/10.1038/nature11446}}

\end{document}